\documentclass[notoc]{JHEP3}  
\usepackage{amsmath,epsfig}
\def\Tr{\mathop{\rm Tr}\nolimits}
\newcommand{\Fslash}[1]{{#1} \!\!\! / }

\author{Daisuke Nomura\\
Theory Center, KEK, Tsukuba, Ibaraki 305-0801, Japan.\\
E-mail: \email{dnomura@post.kek.jp}}

\abstract{
Motivated by the possibility that the right-handed top-quark ($t_R$)
is composite, we discuss the effects of dimension-six
operators on the Higgs boson production at the LHC.  
When $t_R$ is the only composite particle among the
Standard Model (SM) particles, the $(V+A)\otimes (V+A)$ type
four-top-quark contact interaction is expected to have the
largest coefficient among the dimension-six operators, according to
the Naive Dimensional Analysis (NDA).  We find that, to lowest order
in QCD and other SM interactions, the cross section of the SM
Higgs boson production via gluon fusion does not receive corrections
from one insertion of the new contact interaction vertex.  We also
discuss the effects of other dimension-six operators whose
coefficients are expected to be the second and the third
largest from NDA.  
We find that the operator which consists of two $t_R$'s
and two SM Higgs boson doublets can recognizably
change the Higgs boson production cross section
from the SM prediction if the cut-off scale is $\sim 1$TeV.}

\title{Effects of Top-quark Compositeness
       on Higgs Boson Production at the LHC}

\preprint{KEK-TH-1311}


\begin{document}

\section{Introduction}

Even though we know that the Standard Model (SM) describes
physics up to the weak scale very well, 
the mechanism of the electroweak symmetry breaking (EWSB)
still remains unknown.  To improve this situation,
it is very important to search for the Higgs boson
and study its properties in detail since in the SM 
it is the Higgs boson that is responsible for the EWSB.
In view of this, the Large Hadron 
Collider (LHC) provides an excellent opportunity
since it is expected to copiously produce the Higgs boson.

Another interesting particle of the SM is the top-quark.  
Since it is the only known quark whose mass is around the weak scale,
it is natural to speculate that it plays a special role in the EWSB
and/or that it has properties different from those of the other 
quarks.  One of such interesting scenarios is the composite top-quark.
For instance, in the scenario proposed in
Refs.~\cite{Suzuki:1991kh,Lebed:1991qv},
the right-handed top-quark $t_R$ is composite, which 
gives the same low energy predictions for the top-quark
mass and the Higgs boson mass as the $t\bar{t}$
condensation models~\cite{Nambu, Miransky:1988xi, Miransky:1989ds}.
Also from different motivations,
there are increasing interests in the possibility that the
top-quark is composite~\cite{Georgi:1994ha, Agashe:2005vg, Lillie:2007hd,
Pomarol:2008bh, Kumar:2009vs}.  It is therefore important
to study phenomenological consequences from the compositeness
of the top-quark.

If the top-quark is composite, the effects of the
compositeness at low energies are described by higher dimensional
operators, which are suppressed by the composite scale
$\Lambda$~\cite{Georgi:1994ha, Eichten:1983hw}.  
For the left-handed top-quark, $\Lambda$
is constrained to be above the order of a few TeV from
$Z\to b\bar{b}$ decays~\cite{Georgi:1994ha,Alcaraz:2006mx},
while for the right-handed top, the constraint is
weaker: Ref.~\cite{Lillie:2007hd} quotes the bound
$\Lambda/g_{\rm new} \gtrsim$ 80 GeV
from the inclusive top pair production cross section
at the Tevatron, where $g_{\rm new}$
is the effective strong coupling constant which is
associated with the four-top-quark contact interactions.

In this article, we study the effects of the right-handed 
top-quark compositeness on the Higgs boson production at the LHC.
At the LHC, the dominant production process of the SM Higgs
boson is gluon fusion, $gg\to H$, where the 
main contribution comes from the top-quark loop diagrams.
This means that the properties of the top-quark, such as the 
anomalous couplings with gluon or the Higgs boson, directly
affects the cross section of gluon fusion.
In this article, we parametrize the effects of the $t_R$
compositeness
by higher dimensional operators, and study those effects
on $gg\to H$ without assuming a particular new physics model which
makes $t_R$ composite.

This article is organized as following.
In the next section, we set out our framework.
We work with the low-energy effective Lagrangian,
and use the Naive Dimensional Analysis (NDA)~\cite{Manohar:1983md}
to estimate the coefficients of the dimension-six operators
relevant to our analysis.
In Section \ref{sec:tR4}, we discuss the 
effect of the $(V+A) \otimes (V+A)$ type four-top-quark contact
interaction, which is expected to have the largest coefficient
among the dimension-six operators according to NDA.
In Section \ref{sec:newdim6}, we study the 
effects of other dimension-six operators whose coefficients
are expected to be the second and the third largest from NDA.
In Section \ref{sec:conclusions}, we conclude our study.

\section{Effective Lagrangian}
\label{sec:Leff}

When the characteristic scale $\Lambda$ of new physics is high
enough, we can describe the effects of
the new physics at low energies in terms of higher dimensional operators.  
In such a case, we may write the low-energy effective
Lagrangian ${\cal L}_{\rm eff}$ as
\begin{align}
 {\cal L}_{\rm eff} =& {\cal L}_{\rm SM} 
+ \sum_{n \ge 5}\sum_{i} \frac{C_i}{\Lambda^{n-4}} {\cal O}^{(n)}_i,
\label{eq:effL_SM}
\end{align} 
where $i$ is the label for the dimension-$n$
operators ${\cal O}_i^{(n)}$ and $C_i$ is the dimensionless
coupling associated with ${\cal O}_i^{(n)}$.  ${\cal L}_{\rm SM}$
is the SM Lagrangian.  In this article we consider
CP-conserving higher-dimensional operators only.

We are interested in the case where, among the
SM fields, only the right-handed top-quark $t_R$ is composite.
We do not specify an underlying physics which makes $t_R$
composite, and work with
the effective Lagrangian Eq.~(\ref{eq:effL_SM}).

To estimate the coefficients of the higher-dimensional operators,
we use the Naive Dimensional Analysis (NDA)~\cite{Manohar:1983md}.
According to NDA, in the case where only $t_R$ is composite
among the SM particles, the following four-fermion operator
is expected to have the largest coefficient
among dimension-six operators:
\begin{equation}
 {\cal O}_{tt} = \frac12
  \big( \overline{t}^\alpha \gamma_\mu P_R t_{\alpha} \big)
  \big( \overline{t}^\beta \gamma^\mu P_R t_{\beta} \big),
\label{eq:newvertex}
\end{equation}
where $\alpha$ and $\beta$ are color indices\footnote{
\label{footnote1}
We do not consider the operator
$  \big( \overline{t}^\alpha \gamma_\mu P_R
        (T^a)^{~~\beta}_{\alpha} t_{\beta} \big)
  \big( \overline{t}^\gamma \gamma^\mu P_R
        (T^a)^{~~\delta}_{\gamma} t_{\delta} \big)
$
since one can show that 
$
  \big( \overline{t}^\alpha \gamma_\mu P_R
        (T^a)^{~~\beta}_{\alpha} t_{\beta} \big)
  \big( \overline{t}^\gamma \gamma^\mu P_R
        (T^a)^{~~\delta}_{\gamma} t_{\delta} \big)
= (1/3)
  \big( \overline{t}^\alpha \gamma_\mu P_R t_{\alpha} \big)
  \big( \overline{t}^\beta \gamma^\mu P_R t_{\beta} \big)
$
with the help of the Fierz rearrangement and the identity
$(T^a)^{~~\beta}_{\alpha}(T^a)^{~~\delta}_{\gamma}
=  \frac12  \delta_\alpha^\delta \delta_\gamma^\beta
 - \frac16  \delta_\alpha^\beta \delta_\gamma^\delta$,
where $T^a$ ($a=1,\ldots,8$) are the generators of the 
color $SU(3)_C$.}.
$P_R$ is the right-handed projector, 
$P_R \equiv (1+\gamma_5)/2$.  
We call this operator the leading order (LO) in NDA.
According to NDA, the operator (\ref{eq:newvertex})
is expected to have the coefficient 
\begin{align}
 C_{tt} = & c_{tt} g^2_{\rm new},
\end{align}
where $g_{\rm new}$ is the effective coupling constant which is
associated with the new strong interactions which makes $t_R$
composite, and $c_{tt}$ is a constant of the order of one.
The NDA argument requires that $g_{\rm new} \lesssim 4\pi$.
For later discussions, we note here that since ${\cal O}_{tt}$
is Hermitian, the Hermiticity of the Lagrangian implies that $C_{tt}$
must be real.  We also note that ${\cal O}_{tt}$ is even under CP.

We also consider dimension-six operators whose coefficients
are next-to-leading (NLO) and next-to-next-to-leading (NNLO)
in the sense of NDA.  
The NLO operators are the dimension-six operators which have
three $t_R$ and a different up-type quark, namely,
\begin{align}
  {\cal O}_{t_R^3a} \equiv& 
 \left( \overline{t_R}^\alpha \gamma^\mu  t_{R\alpha} \right)
 \left( \overline{t_R}^\beta  \gamma_\mu  u_{R\beta} \right)
+ {\rm h.c.},
\label{eq:OtR3a} \\ 
  {\cal O}_{t_R^3b} \equiv& 
 \left( \overline{t_R}^\alpha \gamma^\mu  t_{R\beta} \right)
 \left( \overline{t_R}^\beta  \gamma_\mu  u_{R\alpha} \right)
+ {\rm h.c.},
\label{eq:OtR3b} \\ 
  {\cal O}_{t_R^3c} \equiv& 
 \left( \overline{t_R}^\alpha (T^a)_{\alpha}^{~~\beta}
             \gamma^\mu  t_{R\beta} \right)
 \left( \overline{t_R}^\gamma (T^a)_{\gamma}^{~~\delta}
             \gamma_\mu  u_{R\delta} \right) 
+ {\rm h.c.},
\label{eq:OtR3c} \\
  {\cal O}_{t_R^3d} \equiv& 
 \left( \overline{t_R}^\alpha (T^a)_{\alpha}^{~~\beta}
             \gamma^\mu  t_{R\delta} \right)
 \left( \overline{t_R}^\gamma (T^a)_{\gamma}^{~~\delta}
             \gamma_\mu  u_{R\beta} \right)
+ {\rm h.c.},
\label{eq:OtR3d} 
\end{align}
and also those operators which are obtained by replacing
$u_R$ in the above operators with $c_R$.  
The NNLO dimension-six operators are those which have exactly two
$t_R$ fields.  Those operators which give corrections
to $gg\to H$ at one-loop are the following:
\begin{align}
 {\cal O}_{tG} \equiv&
 g_s \left[ \bar{t}_R \gamma^\mu T^a D^\nu t_R
      + \overline{D^\nu t_R} \gamma^\mu T^a t_R 
 \right] G^a_{\mu\nu}, \\
{\cal O}_{t4} \equiv & i
\left( \Phi^\dagger \Phi - \frac{v^2}{2} \right)
  \left( \overline{t}_R \gamma^\mu D_\mu t_R 
       - \overline{D_\mu t_R} \gamma^\mu t_R 
  \right),
\end{align}
where $\Phi$ is the SM Higgs boson doublet, $v$ is the vacuum
expectation value (VEV) of the Higgs boson, $v\sim 246$GeV,
and $g_s$ is the gauge coupling constant of QCD.
According to NDA, the NLO and NNLO operators are expected 
to have the coefficients $g_{\rm new}/\Lambda^2$ and 
$1/\Lambda^2$, respectively, up to a coefficient of the order
of one.

There are also four-fermion NNLO operators which consist of
two $t_R$ fields and two other fermions.  These operators
could be constrained from low-energy flavor/precision physics,
and for simplicity, we do not consider these four-fermion NNLO
operators in this article.

The experimental constraints on $\Lambda$ are discussed
in Refs.~\cite{Georgi:1994ha,Lillie:2007hd}.  
In Ref.~\cite{Lillie:2007hd}, from the top-quark pair
production cross section at the Tevatron,
they quote the bound $\Lambda/g_{\rm new} \gtrsim 80$GeV. 
If the NDA estimate is saturated, namely when $g_{\rm new}\sim 4\pi$,
we have $\Lambda \gtrsim 1$TeV.

\section{Corrections to $gg\to H$
from the four-$t_R$ contact interaction}
\label{sec:tR4}

We now discuss the effect of the four-$t_R$ contact interaction
Eq.~(\ref{eq:newvertex}), which is expected to be the 
leading dimension-six operator from NDA, on the gluon fusion
$gg\to H$ using the effective Lagrangian Eq.~(\ref{eq:effL_SM}).
The relevant diagrams are shown in Fig.~\ref{fig:diagrams}.
Interestingly, all the diagrams turn out to vanish separately.
In the following few paragraphs, we discuss why this happens.

\EPSFIGURE{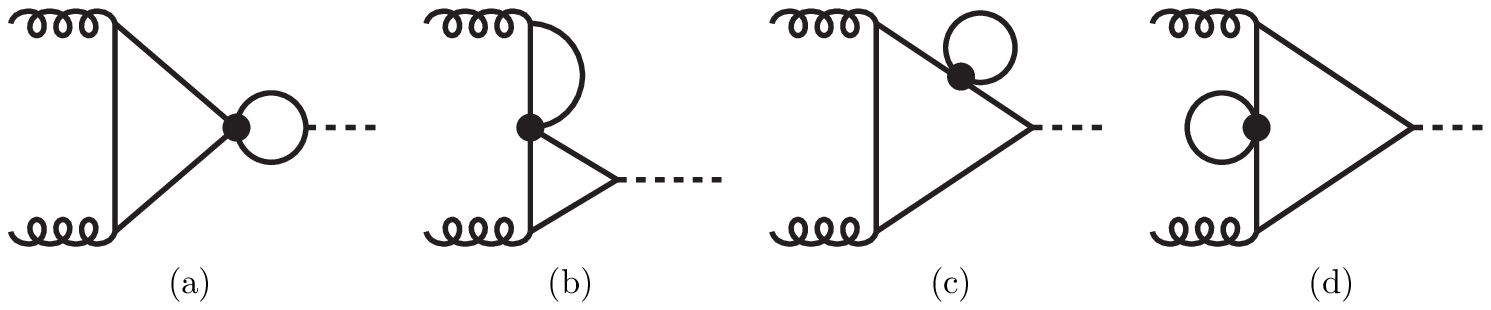}
{\label{fig:diagrams}
Feynman diagrams for the corrections to $gg\to H$
with one insertion of the four-$t_R$ contact interaction vertex,
which is denoted by the black blobs.
The solid lines stand for the top-quark.  The curly and the
dotted lines correspond to the gluon and the Higgs boson,
respectively.}

First let us discuss the diagram Fig.~\ref{fig:diagrams} (a).
By using the Fierz identity where necessary, we see that this
diagram is proportional to the factor,
\begin{align}
 \int \frac{d^d k}{(2\pi)^d}
  \Tr \left[
 \frac{\Fslash{k}+m_t}{k^2-m_t^2}
 \frac{\Fslash{k}+\Fslash{q}+m_t}{(k+q)^2-m_t^2} \gamma^\sigma P_R
   \right],
\label{eq:integral}
\end{align}
where $q$ is the momentum of the Higgs boson and
$d$ is the spacetime dimension.
The integral~(\ref{eq:integral}) comes
from the loop in the subdiagram Fig.~\ref{fig:subdiagrams} (a)
of the diagram Fig.~\ref{fig:diagrams} (a).
It is straightforward to see that this integral
vanishes, by explicitly performing the integral.

\EPSFIGURE{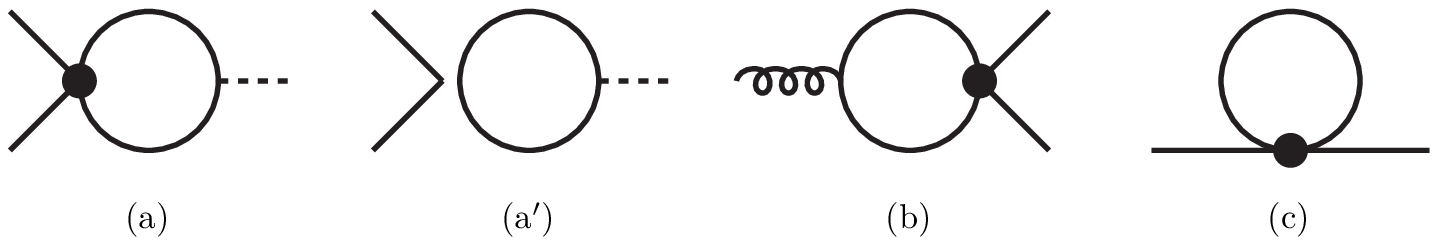}{
\label{fig:subdiagrams}
Some subdiagrams of the diagrams
in Fig.~\ref{fig:diagrams}:
(a) vertex corrections to the $t$-$t$-Higgs interaction.
(a$'$) the ``factorized form'' of the diagram (a), which is
obtained from (a) by using the Fierz rearrangement where 
necessary.  (b) vertex corrections to the gluon-$t$-$t$ interaction. 
(c) corrections to the top-quark self-energy.}

The fact that the diagram Fig.~\ref{fig:diagrams} (a) vanishes
can be understood from a more
general argument.  
First, we should note that, with the help of the Fierz rearrangement,
the subdiagram Fig.~\ref{fig:subdiagrams} (a) of the diagram 
Fig.~\ref{fig:diagrams} (a)
can be written in a ``factorized form'' in
Fig.~\ref{fig:subdiagrams} (a$'$).  Here, by ``factorized form'' 
we mean that the two top-quark fields in one of the
$\bar{t} \gamma^\mu t_R$ factors in ${\cal O}_{tt}$ are
both associated with the external top-quark lines in 
Fig.~\ref{fig:subdiagrams} (a$'$), and the two top-quark fields
in the other $\bar{t} \gamma^\mu t_R$ in ${\cal O}_{tt}$
are contracted with the
top quarks from the top-Yukawa coupling to form the loop.
Next, we should note that if we integrate out high-momentum
modes in the loop of Fig.~\ref{fig:subdiagrams} (a$'$),
we expect that at low energies, the effect of this diagram is
equivalent to a dimension-six operator which consists of
one factor of $\bar{t} \gamma^\mu t_R$ and two SM Higgs boson
doublets.  (Once we require that two $t_R$'s and a SM Higgs boson
doublet should be contained in a dimension-six operator,
we need one more SM Higgs doublet for the
operator to be invariant under the SM gauge group.)
In addition, to contract over the Lorentz 
index $\mu$, we need to include a (covariant) derivative in such a
dimension-six operator.  Summing up the above, the possible candidates
for the dimension-six operator which effectively describes
the diagram Fig.~\ref{fig:subdiagrams} (a$'$) at low energies are:
\begin{align}
 & \big( \overline{D_\mu t} \gamma^\mu t_R \big) \Phi^\dagger \Phi,
\label{eq:DttHH} \\
 & \big( \overline{t} \gamma^\mu D_\mu t_R \big) \Phi^\dagger \Phi,
\label{eq:tDtHH} \\
& \big( \overline{t} \gamma^\mu t_R \big)  (D_\mu \Phi^\dagger) \Phi, 
\label{eq:ttDHH} \\
& \big( \overline{t} \gamma^\mu t_R \big)  \Phi^\dagger D_\mu \Phi, 
\label{eq:ttHDH} 
\end{align}
and their linear combinations.  Now, if we carefully look
into the structure of the diagram Fig.~\ref{fig:subdiagrams} (a$'$),
we see that it is sufficient to consider only the operators
Eqs. (\ref{eq:ttDHH}, \ref{eq:ttHDH}) and their linear combinations.
This is because
the only momentum which can appear in the effective vertex obtained
after integrating over the loop momentum in 
Fig.~\ref{fig:subdiagrams} (a$'$) is that of the Higgs boson.
At this stage, we are left with only two candidates:
\begin{align}
 {\cal O}_{t2} \equiv &
  i \big( \overline{t} \gamma^\mu t_R
    \big)
   \big[ \Phi^\dagger D_\mu \Phi - (D_\mu \Phi)^\dagger \Phi \big], 
\label{eq:Ot2} \\
 \overline{{\cal O}}_{t2} \equiv &
   \big( \overline{t} \gamma^\mu t_R
    \big)
   \big[ \Phi^\dagger D_\mu \Phi + (D_\mu \Phi)^\dagger \Phi \big],
\label{eq:Ot2bar}
\end{align}
where we have taken linear combinations of
Eqs.~(\ref{eq:ttDHH}, \ref{eq:ttHDH}).  The advantage of the
basis Eqs.~(\ref{eq:Ot2}, \ref{eq:Ot2bar}) is that ${\cal O}_{t2}$
and $\overline{{\cal O}}_{t2}$
are Hermitian, and  at the same time, eigenstates of CP:
The operator ${\cal O}_{t2}$ is even under CP, while 
$\overline{{\cal O}}_{t2}$ is CP-odd\footnote{
The effects of these operators at colliders are discussed e.g.\ in
Refs.~\cite{Whisnant:1997qu,Yang:1997iv}.}.  We should also note here
that since both ${\cal O}_{t2}$ and $\overline{{\cal O}}_{t2}$
are Hermitian, their coefficients
in the effective Lagrangian must be real.   Now, 
as discussed in Section 2, since the 
operator ${\cal O}_{tt}$ as well as its coefficient $C_{tt}$
and the top-Yukawa coupling cannot be a source of CP-violation,
the effective vertex for
Fig.~\ref{fig:subdiagrams} (a$'$) must be described by a CP-conserving
operator.  In our case, the only candidate is ${\cal O}_{t2}$, 
but actually this is impossible since after the EWSB,
${\cal O}_{t2}$ reduces to 
$(g_{\scriptscriptstyle Z}/2) (H+v)^2 Z^\mu \left( \bar{t}_R
\gamma_\mu t_R \right)$
in the unitarity gauge, where $g_{\scriptscriptstyle Z}$ 
is the SM gauge coupling associated with the $Z$-boson,
and it does not provide an effective $t_R$-$t_R$-$H$ vertex.
We are now left with no candidate, which means that
the subdiagram Fig.~\ref{fig:subdiagrams} (a$'$) vanishes
after integrating over the loop momentum.
In fact, by explicitly performing the integral
Eq.~(\ref{eq:integral}), we see that it really does.

We now discuss the diagram Fig.~\ref{fig:diagrams} (b).  
This diagram contains
the correction to the gluon-$t_R$-$t_R$ vertex 
as a subdiagram, which we show in Fig.~\ref{fig:subdiagrams} (b).
By judiciously using the Fierz identity when calculating the diagram
Fig.~\ref{fig:diagrams} (b), the subdiagram
Fig.~\ref{fig:subdiagrams} (b) can be shown to be proportional to 
\begin{align}
 \int \frac{d^d k}{(2\pi)^d}
  \Tr \left[
 \frac{\Fslash{k}+m_t}{k^2-m_t^2}\gamma^\mu
 \frac{\Fslash{k}+\Fslash{p}_1+m_t}{(k+p_1)^2-m_t^2} \gamma^\nu P_R
   \right],
\end{align}
where $p_1^\mu$ is the gluon momentum.  This integral is familiar
from the one-loop corrections in QED:  It can be written in the
form, 
\begin{align}
 (p_1^\mu p_1^\nu - g^{\mu\nu}p_1^2) \Pi(p_1^2),
\end{align} 
where $\Pi(p_1^2)$ is a function which does not have a pole at
$p_1^2=0$.
In this expression, the first term vanishes when multiplied by
the gluon polarization vector $\epsilon_\mu(p_1)$, and the
second term also vanishes for on-shell gluons.  It follows that
the diagram Fig.~\ref{fig:diagrams} (b) vanishes for transversely
polarized on-shell gluons.

The diagrams Fig.~\ref{fig:diagrams} (c) and Fig.~\ref{fig:diagrams} (d)
also vanish.  To see this, we look
into the subdiagram shown in Fig.~\ref{fig:subdiagrams} (c).
By using the Fierz identity where necessary, the self-energy
diagram Fig.~\ref{fig:subdiagrams} (c) contains the factor,
\begin{align}
 \int \frac{d^d k}{(2\pi)^d}
  \Tr \left[
 \frac{\Fslash{k}+m_t}{k^2-m_t^2}\gamma^\sigma P_R
   \right],
\end{align}
which vanishes from the $\gamma$ matrix algebra in $d$-dimensions
and from the angular average over $k^\mu$.  Therefore, the diagrams
Fig.~\ref{fig:diagrams} (c) and Fig.~\ref{fig:diagrams} (d) vanish.

Summing up, to lowest order in QCD and other SM interactions,
the parton-level cross section of $gg\to H$ does not receive
corrections from one insertion of the 
contact interaction Eq.~(\ref{eq:newvertex}).

An obvious corollary of this conclusion is that, 
to lowest order in QCD and other SM interactions,
the decay rate of $H \to \gamma\gamma$ does not receive
corrections from one insertion of the 
contact interaction Eq.~(\ref{eq:newvertex}).

\section{Effects from subleading dimension-six operators}
\label{sec:newdim6}

In this section we discuss the effects from subleading
dimension-six operators
on the Higgs boson production at the LHC.

At NLO in NDA, we have the operators
Eqs.~(\ref{eq:OtR3a})-(\ref{eq:OtR3d})
and also those operators which are obtained by replacing
$u_R$ in Eqs.~(\ref{eq:OtR3a})-(\ref{eq:OtR3d}) with $c_R$.
Interestingly, only one of
the operators (\ref{eq:OtR3a})-(\ref{eq:OtR3d})
is independent.  In fact, by using the Fierz transformation
and the identity for the $SU(3)_C$ generators $T^a$
which we mentioned in Footnote~\ref{footnote1}, we can show
that
\begin{align}
  {\cal O}_{t_R^3b} = & {\cal O}_{t_R^3a}, \\
  {\cal O}_{t_R^3c} = & \frac13 {\cal O}_{t_R^3a}, \\
  {\cal O}_{t_R^3d} = & {\cal O}_{t_R^3c}.
\end{align}
Hence it is sufficient to consider the effects from
${\cal O}_{t_R^3a}$.

The effects from the operator ${\cal O}_{t_R^3a}$ on $gg\to H$
is not very interesting:  It gives contribution
only at two-loop order or higher, and hence its effect
on $gg\to H$ is too small to be interesting.

Now we discuss the effects from the NNLO dimension-six operators.
Those NNLO operators which are relevant to $gg\to H$ at one-loop
are ${\cal O}_{tG}$ and ${\cal O}_{t4}$, as discussed
in Section~\ref{sec:Leff}.

The operator ${\cal O}_{tG}$ does not give contribution
to $gg\to H$ at one-loop order.  This can be seen in the
following way.  The operator ${\cal O}_{tG}$ can be
written as
\begin{align}
 {\cal O}_{tG} = & {\rm (total~derivative)}
 - g_s \left[ \bar{t}_R \gamma^\mu T^a t_R \right] D^\nu G^a_{\mu\nu},
\end{align} 
where the first term is a total derivative, which 
does not give contribution to $gg\to H$ as long as we
work in perturbation theory.  The second term can potentially
give contribution to $gg\to H$ through the two diagrams 
in Fig.~\ref{fig:OtG}.  Interestingly, the contributions from
the two diagrams separately vanish:
\EPSFIGURE{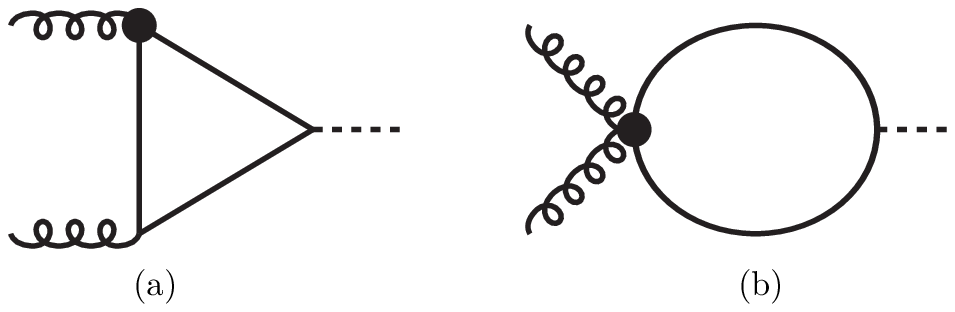}
{\label{fig:OtG}
The contributions from ${\cal O}_{tG}$
to $gg\to H$ at one-loop, which turn out to be zero.}
The diagram Fig.~\ref{fig:OtG} (a) vanishes for the
on-shell gluons since this diagram is proportional to
$\epsilon_\nu(p) p^2 - p^\mu p_\nu \epsilon_\mu(p)$, where
$p^\mu$ is the momentum of the gluon and $\epsilon^\mu(p)$
is the polarization vector of the gluon\footnote{
If we consider the effect of initial state radiation,
the gluons are not necessarily on-shell, and this argument
does not necessarily hold.  Such a process is part of 
the NLO QCD corrections to the Higgs boson production, and
we can give a crude estimate for the size of the QCD corrections as
\begin{align}
 \frac{\sigma({\rm NLO};pp \to H + X)_{{\rm SM}+{{\cal O}_{tG}}}}
      {\sigma({\rm LO}; pp \to H + X)_{\rm SM}}
 \sim& 
 \frac{\sigma({\rm NLO}; pp \to H + X)_{\rm SM}}
      {\sigma({\rm LO}; pp \to H + X)_{\rm SM}}
   \frac{m_H^2}{\Lambda^2}
 \sim {\cal O}(1) \times \frac{m_H^2}{\Lambda^2},
\end{align}
where $\sigma({\rm LO};pp \to H + X)_{\rm SM}$ and
$\sigma({\rm NLO};pp \to H + X)_{\rm SM}$ are the LO and NLO (in QCD)
SM predictions for the Higgs boson production cross section in
$pp$ collisions, respectively, and
$\sigma({\rm NLO};pp \to H + X)_{{\rm SM}+{{\cal O}_{tG}}}$
is the NLO (in QCD) cross section for the same process with
one insertion of ${\cal O}_{tG}$.  In the equation above,
we have included a factor of $1/\Lambda^2$ since the process
involves an insertion of the dimension-six operator, and,
to match the dimension, included a factor of $m_H^2$
since $m_H$ is the typical energy scale of the process.
We see that for $(m_H, \Lambda) \simeq (100{\rm GeV}, 1000{\rm GeV})$
the correction to the cross section is of the order
of 1\% compared to the LO SM prediction.
Of course, a more elaborate calculation is necessary
to obtain a more accurate prediction, which is, however,
beyond the scope of this paper.}. 
The diagram Fig.~\ref{fig:OtG} (b) also vanishes since 
the color factor associated with the top-quark loop
is $\Tr(T^a)=0$.

\EPSFIGURE{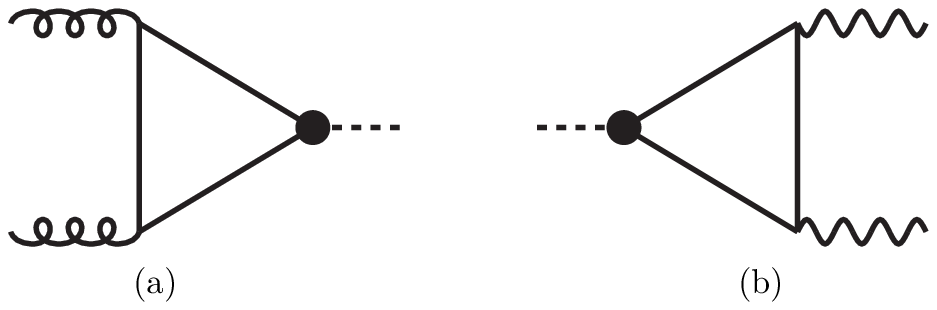}
{\label{fig:Ot4}
(a) The one-loop contribution from ${\cal O}_{t4}$
to $gg\to H$.  (b) The one-loop contribution from ${\cal O}_{t4}$
to $H \to \gamma\gamma$ decay.}

Finally, the operator ${\cal O}_{t4}$ can give finite
correction to $gg\to H$ via the diagram Fig.~\ref{fig:Ot4} (a). 
After a straightforward calculation,
the amplitude ${\cal M}_{t4}$ for Fig.~\ref{fig:Ot4} (a)
turns out to be
\begin{align}
 {\cal M}_{t4} =&
- \frac{C_{t4} v^2}{\Lambda^2} {\cal M}_{\rm SM(top, 1\mbox{-}loop)},
\end{align}
where
${\cal M}_{\rm SM(top, 1\mbox{-}loop)}$ is the LO
SM contribution from the top-quark loop.
The amplitude $ {\cal M}_{t4}$ interferes with the contribution
from the SM contribution, and in total we obtain
\begin{align}
 {\cal M}_{\rm total}
=& \left( 1 - \frac{C_{t4} v^2}{\Lambda^2}
\right) {\cal M}_{\rm SM(top, 1\mbox{-}loop)},
\end{align}
where we neglect the contribution from the $b$-quark loop
and the higher order corrections.
The cross section is then given by
\begin{align}
 \sigma(gg\to H; {\rm total}) =&
\left( 1 - \frac{C_{t4} v^2}{\Lambda^2}
\right)^2 \sigma(gg\to H; {\rm SM(top, 1\mbox{-}loop)}).
\end{align}
\EPSFIGURE[ht]{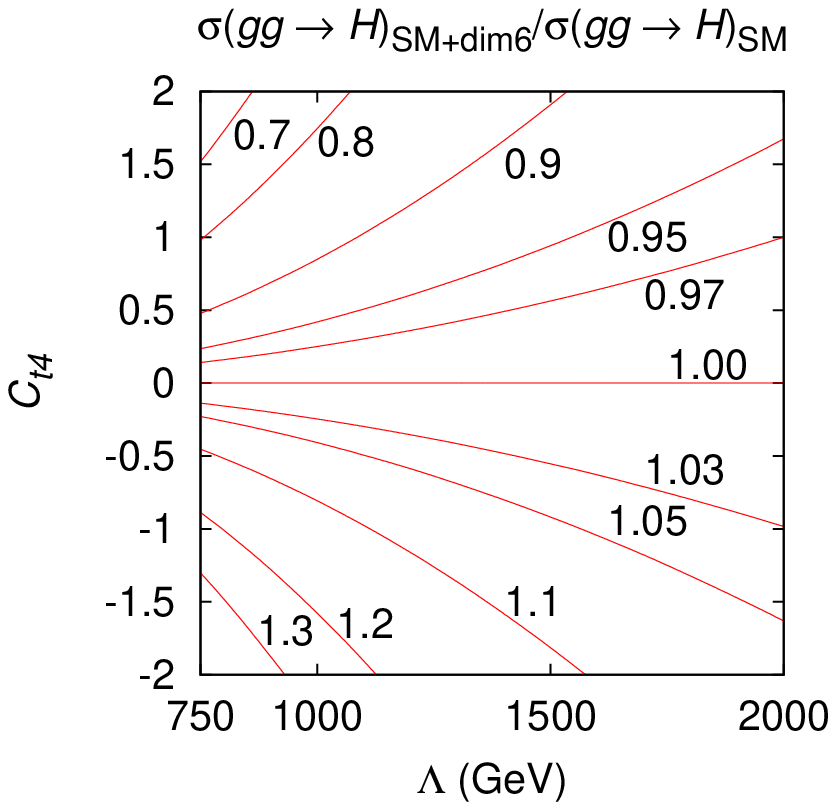}
{\label{fig:ggHratio}
Contour plot of the cross section of $gg\to H$
normalized by the leading order SM prediction
as a function of $\Lambda$ and $C_{t4}$.}
In Fig.~\ref{fig:ggHratio} we show a contour plot of
the cross section of $gg\to H$ normalized by the LO SM
prediction as a function of $\Lambda$ and $C_{t4}$.
In this normalization we expect that the bulk of the QCD
corrections cancel between the numerator and the denominator.
From the figure we see that there are some regions where
the correction can be sizable, for
example, for $(\Lambda, C_{t4}) =(1{\rm TeV}, \pm 2)$,
the correction to the cross section of $gg\to H$
is $\sim \mp 24\%$ compared to the LO prediction in the SM.

The operator ${\cal O}_{t4}$ also gives the correction to the
decay rate of $H\to \gamma\gamma$ by the diagram
Fig.~\ref{fig:Ot4} (b).  For the Higgs mass range
$100{\rm GeV} \lesssim m_H \lesssim 200{\rm GeV}$, this
decay mode is important for the SM Higgs boson searches
at the LHC.  The $H\to \gamma\gamma$ decay width in this case
is given by
\begin{align}
 \Gamma(H \to \gamma\gamma) = &
 \frac{\alpha^2 g^2}{1024\pi^3} \frac{m_H^3}{m_W^2}
 \left|
   N_c Q_t^2 \left( 1 - \frac{C_{t4}v^2}{\Lambda^2}\right)
   F_{1/2} \left(\frac{4m_t^2}{m_H^2}\right)
 +  F_1 \left(\frac{4m_W^2}{m_H^2}\right)
 \right|^2 ,
\end{align}
where $N_c=3$ and $Q_t=2/3$.  In the above expressions we
use the approximation that we
include only the $W$-boson and the $t$-quark contributions.  
The functions $F_{1/2}(\tau)$ and $F_{1}(\tau)$ are given by
\begin{align}
 F_{1/2}(\tau) =& -2\tau [ 1 + ( 1 - \tau) f(\tau)], \\
 F_{1}(\tau) =& 2 + 3 \tau + 3 \tau ( 2 - \tau) f(\tau),\\
 f(\tau) =&
\begin{cases}
[\sin^{-1}(\sqrt{1/\tau})]^2,     
                      & \text{for $\tau \ge 1$,} \\
-\frac14 [\ln( (1+\sqrt{1-\tau})/  (1-\sqrt{1-\tau}))
 - i\pi ]^2 ,
                      & \text{for $\tau < 1$,}
\end{cases}
\end{align}
which are the functions which appear in the
SM contribution to the $H\to \gamma\gamma$ decay
(for review, see e.g.~Ref.~\cite{Gunion:1989we}).
In Fig.~\ref{fig:Hgamgamratio} we show a contour plot of
$\Gamma(H\to\gamma\gamma)$ normalized by the LO SM
prediction as a function of $\Lambda$ and $C_{t4}$. 
In the figure we take $m_H=120$GeV, but this figure
does not change very much for $100{\rm GeV} \lesssim m_H
\lesssim 200$GeV.  We find that the correction
to $\Gamma(H\to \gamma\gamma)$ is opposite in sign 
to the correction to $\sigma(gg\to H)$.
We also see that the correction is smaller than
that to $\sigma(gg\to H)$, and is about $\pm 8\%$ for
$\Lambda =1$TeV and $C_{t4}=\pm 2$.  
In Fig.~\ref{fig:ggHgamgamratio}, we show a contour plot of the 
product of $\sigma(gg\to H)$ and $\Gamma(H\to\gamma\gamma)$
as a function of $\Lambda$ and $C_{t4}$.
Also in this figure we take $m_H=120$GeV, even though this figure
does not change very much for $100{\rm GeV} \lesssim m_H
\lesssim 200{\rm GeV}$.
From Fig.~\ref{fig:ggHgamgamratio}, we see that the correction to 
$\sigma(gg\to H)$ and that to $\Gamma(H\to\gamma\gamma)$
partially cancel with each other, and
we are left with $\pm 16\%$ corrections 
to $\sigma(gg\to H)\Gamma(H\to \gamma\gamma)$
at $(\Lambda, C_{t4}) =(1 {\rm TeV}, \mp 2)$.

\EPSFIGURE[ht]{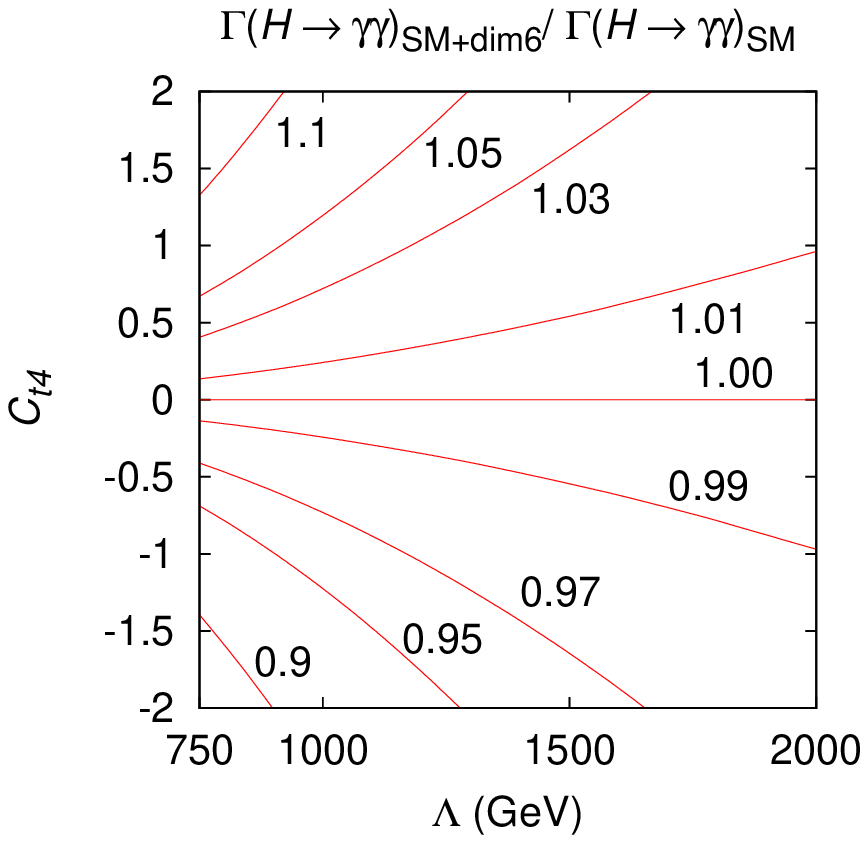}
{\label{fig:Hgamgamratio}
Contour plot of the decay rate of $H \to \gamma\gamma$
normalized by the leading order SM prediction
as a function of $\Lambda$ and $C_{t4}$.}

\EPSFIGURE[ht]{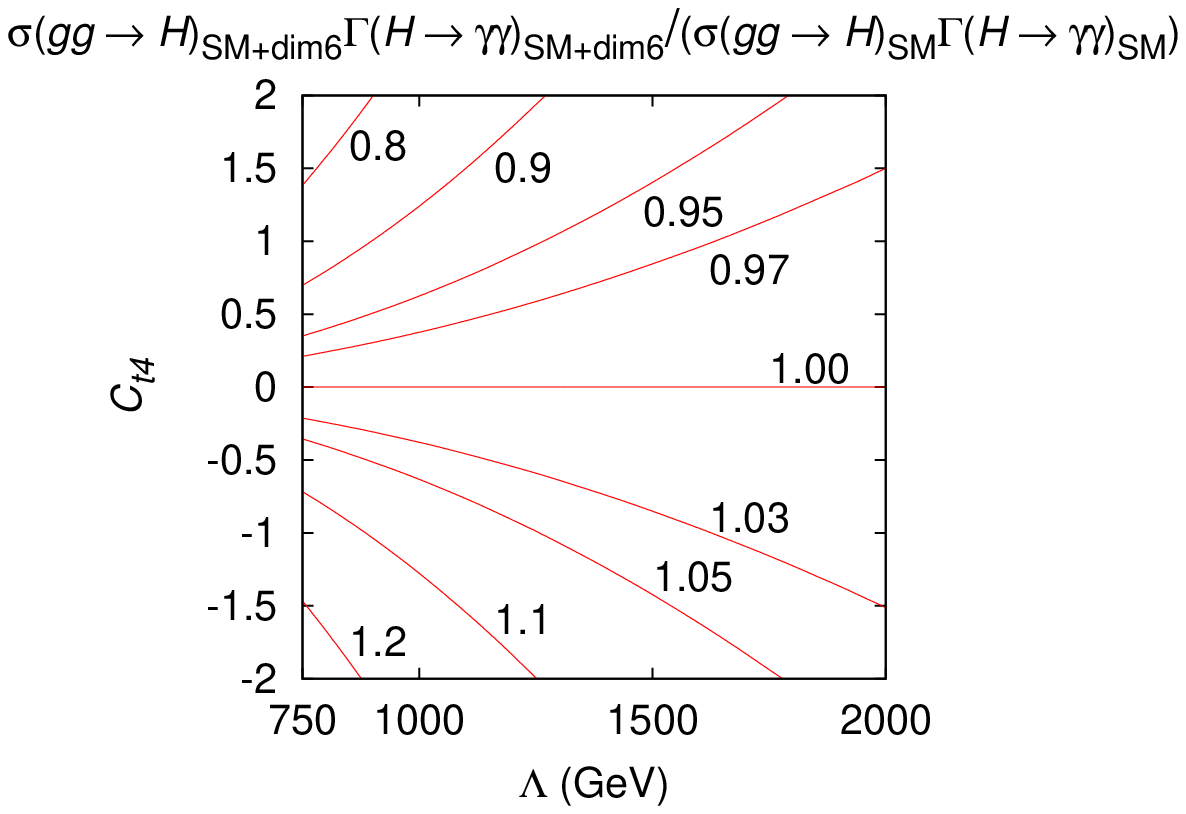}
{\label{fig:ggHgamgamratio}
Contour plot of the cross section of $gg\to H$ times
$\Gamma(H \to \gamma\gamma)$,
normalized by the leading order SM prediction
as a function of $\Lambda$ and $C_{t4}$.}

\section{Conclusions}
\label{sec:conclusions}

We have studied the effects of the right-handed top-quark
compositeness on the Higgs boson production at the LHC.  
We find that, to lowest order in QCD and other SM interactions,
there is no
correction from one insertion of the four-$t_R$ contact interaction,
whose coefficient in ${\cal L}_{\rm eff}$
is expected to be the largest among dimension-six operators,
according to NDA.
We also find that the NLO dimension-six operators (in the sense of NDA)
are four-fermion contact interactions which involve 
three $t_R$'s and a different right-handed up-type quark.
This operator does not give corrections to $gg\to H$
up to and including one-loop.  Finally, we find that 
the NNLO operator ${\cal O}_{t4}$
can give correction
of $\sim \pm 24\%$ to the Higgs boson production cross section
$\sigma(gg\to H)$
for $\Lambda = 1$TeV and $C_{t4}=\mp 2$.  For the same parameters,
we see that the correction to the decay rate of
the Higgs boson to two photons is about $\mp 8\%$,
where the sign is opposite to that of the correction
to $\sigma(gg\to H)$.  In total, for $\Lambda=1$TeV
and $C_{t4}= \mp 2$, we expect $\pm 16\%$ correction in the product of 
the production cross section $\sigma(gg\to H)$ 
and the partial decay rate $\Gamma(H\to \gamma\gamma)$,
compared to the SM prediction,
if the Higgs boson mass is in the range between 100GeV
and 200GeV.  This effect would be recognizable in the Higgs boson
searches at the LHC.  If the Higgs boson is heavier, and 
the $WW$ decay channel is open, then we do not have to rely
on the $H\to \gamma\gamma$ channel.  In this case,
the correction to $\sigma(gg\to H) \Gamma(H\to WW)$ is 
$\sim \pm 24\%$ for $\Lambda = 1$TeV and $C_{t4}=\mp 2$,
which means that the effect from the top-quark compositeness
will be a little bit more clearly seen in this channel.

\acknowledgments
The author would like to thank M.\ Hashimoto and S.\ Kanemura for
useful discussions.
The work of D.N. was supported by the 
Research Fellowship of the Japan Society
for the Promotion of Science for Young Scientists.

\end{document}